# HALO/THICK DISK CVS AND THE COSMIC X-RAY BACKGROUND

*To Appear in the IAU Symp. No. 168*

JONATHAN E. GRINDLAY AND EYAL MAOZ
*Harvard-Smithsonian Center for Astrophysics,*
*60 Garden Street, Cambridge, MA 02138*

**Abstract.** In a recent study (Maoz and Grindlay 1995) we have found that a number of previously recognized anomalies in the diffuse x-ray background at soft energies (~0.5-2 keV) can be understood if about 20-30% of the diffuse flux arises from a population of low luminosity sources in a thick disk or flattened halo distribution in the Galaxy. Here we summarize our results and review the arguments that these objects are not accreting neutron stars or black holes but rather white dwarfs (i.e. CVs) which may have been produced in a primordial population of disrupted globular clusters.

## 1. Introduction

The COBE results (Mather et al 1990) provided final confirmation that the cosmic x-ray background (CXB) radiation is not due to hot gas but, as long suspected (Giacconi et al 1979, Soltan 1991) due to the superposition of discrete sources. It is the nature of these sources that has been the major question. Although many arguments (e.g. the fluctuation analyses of Hamilton and Helfand (1987) and Barcons and Fabian (1988)) point to the dominance of relatively low luminosity active galactic nuclei (AGN), the lack of clustering strongly suggests that in fact no more than about 70% of the CXB can arise from known classes of AGNs. A new population of weak sources is probably required. Thus it was extremely interesting when in an epochal study of the CXB and source counts with the Deep Survey observations undertaken with ROSAT, Hasinger et al (1993; hereafter H93) concluded that in fact about 60% of the CXB at fluxes $\geq 2.5\times10^{-15}$ cgs (hereafter cgs = erg cm$^{-2}$ s$^{-1}$ ) in the 0.5-2 keV band could indeed be resolved into discrete sources but that the source counts are inconsistent with models for the x-ray luminosity functions (XLF) of AGNs (Boyle et al 1993). H93 concluded, therefore, that either more complicated models for the evolution of the XLF or – again– that a new population of sources is needed at the faintest flux levels.

It is this possibility that we have explored (Maoz and Grindlay 1995, hereafter MG). In this brief addendum to that work, we summarize our analysis and results for constraints on the possible luminosity and density of a *galactic* population of faint x-ray sources (halo/thick disk CVs) which could satisfy all the new constraints from ROSAT as well as the earlier CXB flux and fluctuation constraints.

## 2. Generalized Halo/Thick Disk Models

We have considered whether either a spherical halo ($\propto 1/R^2$) or a thick disk (exponential in both R and Z) distribution of discrete sources of fixed (standard candle) luminosity $L$ and spatial density $n_0$ (in the solar neighborhood) can satisfy all the constraints imposed by the source counts in the 27 deep fields surveyed by H93 as well as the fluctuation and flux constraints imposed by the CXB. Our models are normalized to a possible galactic contribution to the source counts of 120 deg$^{-2}$, at the flux limits of H93, which allows for the measured contributions of stars and all known classes of extragalactic sources (galaxies, clusters, BL Lacs and QSOs with the XLF evolution of Boyle et al 1993). Details of the calculation are given in MG and our results are shown below in Figure 1, which presents all the constraints on the allowed values of L and $n_0$ for two spherical halo models with small and large core radii, and for two thick disk models with modest and large scale heights. For each model, the solid line represents the combinations of $L$ and $n_0$ that would exactly resolve the discrepancy found in the faint source counts; the shaded area corresponds to combinations which would produce a number density of unidentified bright sources which is consistent with observations; the two inclined dashed lines confine the parameter space which would not violate constraints on the contribution to the CXB intensity; the horizontal dash-dot lines confine the region which would be consistent with the results of fluctuation analyses of the unresolved background, and the region below the dotted line corresponds to models which contribute $\leq 10^{40}$erg/s to the Galaxy's total x-ray luminosity.

The striking result is that for either halo or thick disk models a relatively well defined region of parameter space is defined by all the constraints. In particular, the typical luminosity of the sources is $10^{31\pm0.5}$erg/s (0.5-2 keV), and is not very sensitive to changes in the scale of the spatial distribution. Thus although Kshyap *et al.* (1994) have recently suggested that MACHOs may have x-ray coronae with typical luminosities of order $\sim 10^{27}$ erg/s, and may produce a small fraction ($\sim$10%) of the soft CXB *flux*, their luminosity is far too low to satisfy the constraints imposed by number counts and the excess of faint sources detected by H93.

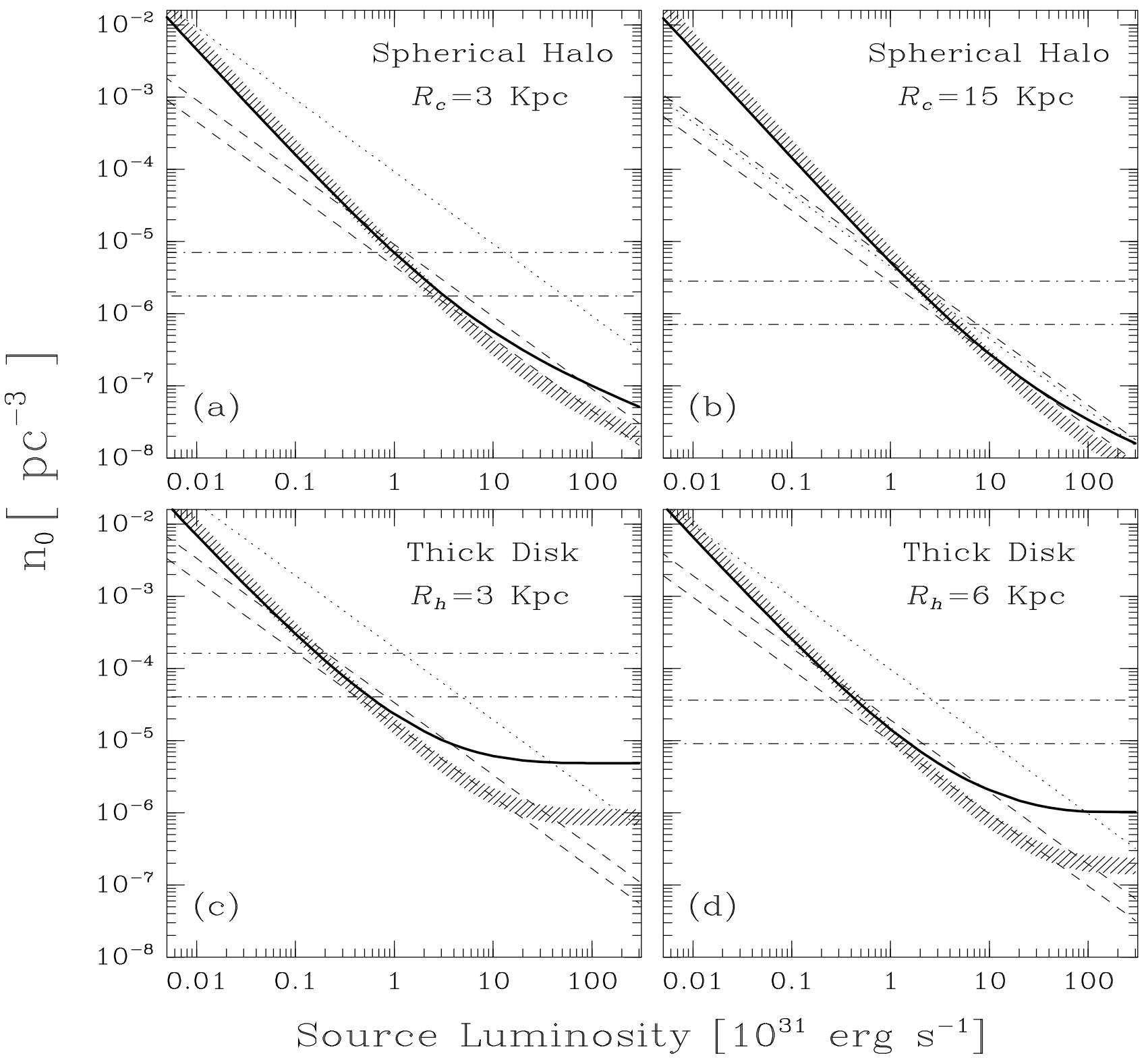

*Figure 1.* Combined observational constraints on the typical x-ray luminosity of the sources, $L$, and their number density in the Solar vicinity, $n_0$, for two different halo models (a,b) and thick disk models (c,d).

All our models have an approximately Euclidean logN-logS relation at fluxes above $\sim 10^{-14}$ cgs, and which flattens considerably below a flux of a few times $10^{-16}$ cgs. In the flux range of $10^{-15}$-$10^{-14}$ cgs, the logN-logS relation is steeper than that expected for an evolving population of QSOs (Boyle *et al.* 1993), thus providing an increasing relative contribution of the Galactic sources with decreasing flux.

We have also calculated the constraints imposed by the spectrum and anisotropy (limits) of the CXB on our proposed source population. H93 found that the average spectrum of the resolved sources becomes harder with decreasing flux. Since the extragalactic sources have a steeper average spectrum in the 0.5 - 2 keV band, we conclude that the average spectral index of the proposed population of sources must be $< 1$ in this band. Allowing for the Galactic emission components of the CXB which already indicated thin *and* thick disk components at higher flux levels, we conclude that the source population proposed here most probably has a flattened distribution and thus should have a significant contribution to the detected anisotropy. The fractional contribution of the proposed population to the CXB in the 2-60 keV band, averaged over the entire sky, is $\sim 8\%$ but can be 20-30% near the galactic center depending on the model. Anisotropy constraints would allow a slightly higher source temperature in more spherical models than in flattened models.

## 3. Identification of the Sources as CVs

Our primary requirements are sources with typical $L_x \sim 10^{31}$ erg/s and a relatively hard spectrum. These are not consistent with dM stars (which are only this bright during brief flares, and are typically soft), with LMXBs in quiescence (the ROSAT studies of Verbunt et al 1994 have shown these sources to be typically very soft, with kT $\sim$ 0.3 keV) or with isolated neutron stars (e.g. old pulsars, for which thermal emission is also expected to be very soft – cf. Ögelman et al 1993). However pulsars also appear to have harder emission components (Ögelman 1994), but with much lower $L_x$ than we require. Magnetized neutron stars (NSs) accreting from the ISM could produce hard cyclotron emission (cf. Nelson et al 1994) with $L_x \sim 0.05 L_{acc} \sim 3 \times 10^{25} n_{-3}/v_{100}^3$ erg/s for a typical pulsar at 100 km/s moving through a halo interstellar medium of density $n_H \sim 10^{-3}$ cm$^{-3}$, which is again far too low. Thus although (old) pulsars would nicely satisfy the requirement for some $10^{8.5}$ total sources in a thick disk or halo, it seems they could not satisfy the luminosity constraint.

Similarly, we now consider $\sim 10^8$ galactic black holes (BHs) with mass $\lesssim 10^4 M_\odot$ (so as not to exceed the halo mass) and moving on halo orbits with v $\sim$ 200 km/s. Note these are probably not the $\sim 10^6 M_\odot$ halo BHs invoked by Lacey and Ostriker (1985) to stabilize the disk although it is conceivable that our lower mass could be accommodated with changes in the primordial fluctuation spectrum (Ostriker, private communication). To obtain the required $L_x$ by accretion from the halo ISM would imply a radiation efficiency $\epsilon \sim 10^{-3}$ ($v_{200}^3/n_{-3}$. However at the low Bondi accretion rates implied ($\sim 10^{-9} n_{-3}/v_{200}^3$ of Eddington), the recent results of Narayan and Yi (1994) for the effective accretion efficiencies onto BHs when advection is included would suggest much lower efficiencies (typically $\lesssim 10^{-5}$, but dependent on the assumed viscosity parameter $\alpha$). Thus BHs are unlikely but perhaps cannot be totally excluded.

Cataclysmic variables (hereafter CVs) can most naturally fit our requirements for luminosity and local space density. Hertz *et al.* (1990), from optical identifications of the x-ray selected Einstein galactic survey (Hertz and Grindlay 1984), find a local space density for low-$\dot{M}$ CVs which is consistent with our prediction. Howell and Szkody (1990) discuss the halo CV population and report that their significantly lower mean orbital period implies that these systems are relatively old. Magnetic CVs (*e.g.*, AM-Her type) are also predominantly below the period gap, and are also optically fainter than nonmagnetic CVs (larger $L_x/L_v$ ratio) since they do not have well developed accretion disks. Thus, it is possible that the proposed population of x-ray sources are some kind of magnetic CVs. Their low-$\dot{M}$ could push their characteristically soft x-ray spectral components into the XUV band leaving just their hard components.

Globular clusters contain dim sources with typical soft x-ray luminosity of $\sim 10^{31.5-32.5}$ erg/s in the ROSAT band, and the bulk of these sources are fully consistent with being CVs (Grindlay 1994a, and references therein). X-ray and optical observations indicate a presence of $\gtrsim 10$ CVs per cluster (at least for (near-) core collapsed globulars) with $L_x \sim 10^{30.5-31.5}$ erg/s. These may be optically fainter than disk CVs for a given x-ray luminosity (Grindlay 1994a), deficient in dwarf novae, and contain a significant fraction of magnetic CVs (Grindlay 1994a,b).

Thus the thick disk or halo CV population proposed in this paper resembles the globular cluster CV population, suggesting they may have been ejected from globular clusters. Indeed, Grindlay (1994b) shows that the dim sources in globulars appear to have an extended radial distribution which extends to radial offsets well beyond what purely relaxed objects (CVs) should obtain and may thus imply ejection. Although this process may contribute to the proposed Galactic population, it obviously cannot provide the required number of $\sim 10^{8.5}$ CVs in the halo from the present $\sim$150 globular clusters. Thus the required halo CVs, if from globulars, must imply a population of some $10^{6-7}$ primordial globulars which have been disrupted and which had a flatter radial distribution than the $\sim 1/R^3$ distribution of the survivors. In particular, a thick disk may form from globular clusters on low-inclination orbits which have been disrupted by encounters with giant molecular clouds, as suggested by Grindlay (1984) for the origin of the field LMXBs.

## 4. Conclusions

The CXB may partially originate in a large population of intrinsically faint (and presumably old) CVs which are distributed within a thick Galactic disk (or an oblate halo) with a scale height of a few kpc. The inferred $\sim 3 \times 10^8$ CVs would have luminosity of $\approx 10^{30-31}$ erg/s and a number density of $\sim 10^{-5}$-$10^{-4}$ pc$^{-3}$ sources in the Solar vicinity. Our model can be tested by searching for a large-scale variation in the surface *number* *density* of resolved x-ray sources. We predict that the density of sources with flux below $10^{-14}$ cgs will decrease either with an increasing galactic latitude ($|b|$) or with an increasing angular distance to the Galactic center ($\theta$). The data presented by H93 are insufficient for performing this test since they do not represent a broad range in either $b$ or $\theta$ and the number of sources is limited. Deeper HRI surveys (for more precise source positions) at several $\theta$ and $|b|$ values are needed.

The most direct test of our CV contribution to the CXB is to optically identify the predicted ~20% of the faintest ROSAT survey sources. Regardless of whether these are non-magnetic CVs or magnetic ones (*e.g.*, AM-Her CVs), they should appear almost as "pure" emission line sources and could thus be found by very deep images in broad band (R) vs. H$\alpha$. We emphasize that *all* faint x-ray sources in the ROSAT fields should be examined, including ones already "identified" as AGNs since it is probably that some fraction of these have been misidentified given an average AGN angular separation of only $\sim 4'$.

We thank Jerry Ostriker, Rosanne Di Stefano, and Elihu Boldt for interesting discussions. This work was supported by the U.S. National Science Foundation, grant PHY-91-06678, and NASA, grant NAGW-3280.